\documentclass[11pt]{article}

\usepackage{a4}
\usepackage{pst-all}
\catcode`\@=11%
\usepackage{graphicx}
\usepackage{subfigure}
\usepackage{natbib}

\psset{unit=9.37mm,arrowscale=1.5}
\SpecialCoor
\usepackage{amssymb}
\usepackage{color}
\usepackage{amsmath}
\usepackage{url}

\newcommand{\BPi}{\mbox{{\boldmath  $\Pi$}}}
\newcommand{\Bsigma}{\mbox{{\boldmath $\Sigma$}}}
\newcommand{\Btheta}{\mbox{{\boldmath $\Theta$}}}
\newcommand{\Bmu}{\mbox{{\boldmath  $\mu$}}}
\newcommand{\Bnu}{\mbox{{\boldmath  $\nu$}}}

\newcommand{\transp}{{\sf T}}

\begin{document}

\title{A BGe score for tied-covariance mixtures \\of Gaussian Bayesian networks}

\date{}
\maketitle
\vspace{-1.5cm}

{\large\centering \textbf{Marco Grzegorczyk}
	\vspace{+0.5cm} \normalsize
	
Bernoulli Institute (BI) \\
Rijksuniversiteit Groningen \\
\emph{E-mail: m.a.grzegorczyk@rug.nl}  \\

\vspace{+0.5cm}}

\begin{abstract}
Mixtures of Gaussian Bayesian networks have previously been studied under full-covariance assumptions, where each mixture component has its own covariance matrix. We propose a mixture model with tied-covariance, in which all components share a common covariance matrix. Our main contribution is the derivation of its marginal likelihood, which remains analytic. Unlike in the full-covariance case, however, the marginal likelihood no longer factorizes into component-specific terms. We refer to the new likelihood as the BGe scoring metric for tied-covariance mixtures of Gaussian Bayesian networks. For model inference, we implement MCMC schemes combining structure MCMC with a fast Gibbs sampler for mixtures, and we empirically compare the tied- and full-covariance mixtures of Gaussian Bayesian networks on simulated and benchmark data.
\end{abstract}

\noindent {\bf Keywords} \\
Gaussian Bayesian networks,  mixture model,  tied-covariance, Markov chain Monte Carlo

\section{Introduction} 
\label{sec:introduction}
Bayesian networks (BNs) are popular graphical models for representing conditional (in-) dependencies among random variables using directed acyclic graphs (DAGs). Each node in a DAG corresponds to a variable, and directed edges encode probabilistic relationships. This structure allows joint distributions with many conditional independencies to be expressed compactly. BNs are widely used for learning under uncertainty, making them important tools in statistics, machine learning, and artificial intelligence. Thorough introductions to Bayesian networks are available in standard textbooks \citep{Pearl, Neapolitan, Koller_Fried, SCUTARI_BOOK}. Structure learning algorithms aim to infer DAGs from data, but since the search space grows super-exponentially with the number of nodes \citep{Chickering_NPhard}, the problem is computationally challenging. For efficient structure learning algorithms, see \cite{Kuipers_2017, scutari2018learns, SCUTARI_2019, SCUTARI_NEERLANDICA, Kitson2023}. In the case of sparse data, there is typically considerable uncertainty about the DAG. This uncertainty can be addressed by Bayesian structure learning algorithms, by sampling DAGs from the posterior distribution and averaging over the sampled DAGs. This approach is commonly referred to as Bayesian model averaging \citep{Constantinou_2022}. For Bayesian BN structure learning algorithms, see \cite{MadiganYork, GiudiciMCMC, FriedKollerOrderMCMC} or \cite{GrzegorczykHusmeierMach08}.  \\

Beyond the standard BN structure learning literature, tailored approaches have been developed for incomplete \citep{Fried_MS-EM, SOFT2, SCUTARI_REASONING}, ordinal \citep{OSEM, OTHER, scauda2025latentcausalinferenceframework}, and hybrid data \citep{HeckermanPrior2, SCUTARI_NEERLANDICA}. We have recently contributed to this line of work by proposing Bayesian counterparts for these important special cases, addressing missing \citep{Gryze2023}, ordinal \citep{Gryze2024}, and hybrid \citep{Gryze2025} data. In the present paper, we extend this trilogy \citep{Gryze2023, Gryze2024, Gryze2025} by introducing a Bayesian approach for tied-covariance mixtures of Gaussian Bayesian networks. A mixture model assumes that the data stem from several subpopulations, each represented by its own distribution. Mixture models are used in many frameworks, and we are not the first to propose mixture Bayesian networks: both non-Bayesian \citep{Ko_2007} and Bayesian \citep{GrzegorczykHusmeierBioinf08} formulations have been introduced, each assuming that all subpopulations share a common DAG but differ in their network parameters. Because they use component-specific covariance matrices, these earlier models are full-covariance mixtures of Gaussian Bayesian networks, which we refer to as M1. The Bayesian M1 model from \cite{GrzegorczykHusmeierBioinf08} employs the structure MCMC sampler \citep{MadiganYork, GiudiciMCMC} to infer the DAGs and the allocation sampler \citep{NobileAllocation} to assign the data to $K$ mixture components. Each mixture component $k=1,\ldots,K$ has a weight $\pi_k\in[0,1]$, a mean vector $\Bmu_k$, and a covariance matrix $\Bsigma_k$. The resulting density is 
$$p({\bf x}|\mbox{DAG},\mathcal{M}_1) = \sum_{k=1}^K \pi_k  \cdot  \mathcal{N}_n({\bf x}|\Bmu_k,\Bsigma_k),  $$
where $\mathcal{N}_n({\bf x}|\Bmu,\Bsigma)$ denotes the density of the $n$-variate normal distribution $\mathcal{N}_n(\Bmu,\Bsigma)$ at ${\bf x}$, and each $\Bsigma_k$ must be consistent with the DAG, that is, each $\Bsigma_k$ must encode the conditional independence relations implied by the DAG. For a dataset $\mathcal{D}=\{{\bf x}_1,\ldots,{\bf x}_m\}$ with given class labels and under suitable prior assumptions,\footnote{In each component $k$, an independent Normal–Wishart prior is used \citep{GeigerHeckGaussUAI, HECKERMAN_ANNALS, HECKERMAN_ADDENDUM}.} the marginal likelihood of the mixture BN model factorizes into the product of the component-specific marginal likelihoods: 
$$ p(\mathcal{D}|\mbox{DAG},\mathcal{M}_1) = \prod_{k=1}^K p(\mathcal{D}_k|\mbox{DAG}), $$
where $\mathcal{D}_k$ denotes the subset of observations with class label $k$, and each $p(\mathcal{D}_k|\mbox{DAG})$ can be computed analytically. \\

In this paper, we introduce the tied-covariance counterpart of the M1 model, which we refer to as the M2 mixture model. In the M2 model the $K$ components share a common covariance matrix $\Bsigma$:
$$p({\bf x}|\mbox{DAG},\mathcal{M}_2) = \sum_{k=1}^K \pi_k  \cdot  \mathcal{N}_n({\bf x}|\Bmu_k,\Bsigma).  $$
The main contribution of this paper is the derivation of the marginal likelihood $p(\mathcal{D}| \text{DAG}, \mathcal{M}_2)$ for the tied-covariance mixture model (M2). We show that it admits an analytic solution, but unlike in the M1 case, it no longer decomposes into $K$ local marginal likelihoods. We refer to the new marginal likelihood as the BGe scoring metric for tied-covariance mixtures of Gaussian Bayesian networks. We implement both types of mixture models and compare their performances on simulated and benchmark data. Our Markov chain Monte Carlo (MCMC) sampling algorithm combines structure MCMC sampling of DAGs \citep{MadiganYork, GiudiciMCMC} with a recently proposed fast Gibbs sampler for mixture models \citep{newman2025fast}.

\section{Statistical methods}
\label{sec:methods}

\subsection{Bayesian networks}
\label{sec:bn}
Bayesian networks (BNs) are graphical models that use directed acyclic graphs (DAGs) to represent conditional independencies among a set of random variables, which we denote by $X_1,\ldots,X_n$. The variables correspond to the $n$ nodes of a DAG, and directed edges between nodes encode direct conditional dependencies. Node $X_j$ is called a parent of $X_i$ if there is an edge $X_j \rightarrow X_i$, and it is called an ancestor of $X_i$ if $X_i$ can be reached from $X_j$ by following a sequence of one or more directed edges along their orientations, symbolically $X_j \rightarrow \ldots \rightarrow X_i$. By definition, a node cannot be its own ancestor, so cyclic paths of directed edges, such as $X_i \rightarrow \ldots \rightarrow X_i$, are excluded. The set of parent nodes of $X_i$ is denoted by $\BPi_{i}$. A DAG can be identified with its $n$ parent sets $\BPi_{1},\ldots,\BPi_n$. Given $\mathcal{G}$, i.e. the conditional independence relations implied by $\mathcal{G}$, the joint density factorizes as
\begin{equation}
    \label{EQ_FACTORIZATION}
    p(X_1,\ldots,X_n|\mathcal{G}) = \prod\limits_{i=1}^n p(X_i|\BPi_i).
\end{equation}
Two DAGs, say $\mathcal{G}_{\star}$ and $\mathcal{G}_{\bullet}$, that imply the same conditional independencies are called equivalent, and they yield the same factorization:
\begin{equation}
\nonumber
p(X_1,\ldots,X_n|\mathcal{G}_{\star}) =   \prod\limits_{i=1}^n p(X_i|\BPi^{\star}_i) = \prod\limits_{i=1}^n p(X_i|\BPi^{\bullet}_i) =
p(X_1,\ldots,X_n|\mathcal{G}_{\bullet}),
\end{equation}
where $\BPi^{\star}_i$ and $\BPi^{\bullet}_i$ denote the parent sets of $X_i$ in $\mathcal{G}_{\star}$ and $\mathcal{G}_{\bullet}$, respectively.
Equivalent DAGs belong to the same equivalence class. \citet{CHICK_2002} shows that DAG equivalence classes can be represented by completed partially directed acyclic graphs (CPDAGs), and that two DAGs $\mathcal{G}_{\star}$ and $\mathcal{G}_{\bullet}$ are equivalent if and only if they share the same skeleton and the same v-structures. The skeleton of a DAG is obtained by replacing each directed edge with an undirected edge. A v-structure is a configuration in which a node $X_i$ has two parent nodes with no edge between them, for example $X_k \rightarrow X_i \leftarrow X_l$ with no edge between $X_k$ and $X_l$.
CPDAGs contain a mixture of directed and undirected edges. A directed edge $X_j \rightarrow X_i$ in a CPDAG indicates that all DAGs in the equivalence class share this directed edge. An undirected edge $X_j - X_i$ indicates that all DAGs in the equivalence class contain an edge between $X_j$ and $X_i$, but not all agree on its orientation. In particular, there is at least one DAG with the edge $X_j \rightarrow X_i$ and at least one DAG with the edge $X_j \leftarrow X_i$. \\

 The random variables $X_1,\ldots,X_n$ can be stacked into a random vector ${\bf X} = (X_1,\ldots,X_n)^{\transp}$. A dataset $\mathcal{D}$ consists of $m$ independent realizations of ${\bf X}$:
$$\mathcal{D} =\{{\bf x}_1,\ldots,{\bf x}_m\}, $$
where ${\bf x}_i$ denotes the $i$-th observation of ${\bf X}$. The posterior distribution of DAG $\mathcal{G}$ is defined as:
\begin{equation}
    \label{EQ_POSTERIOR_DAG}
    p(\mathcal{G}|\mathcal{D}) := \frac{p(\mathcal{D}|\mathcal{G})p(\mathcal{G})}{p(\mathcal{D})} \propto p(\mathcal{D}|\mathcal{G})p(\mathcal{G}),
\end{equation}
where $p(\mathcal{D}|\mathcal{G})$ is the marginal likelihood of $\mathcal{G}$,  $p(\mathcal{G})$ is the prior probability of $\mathcal{G}$, and 
$p(\mathcal{D})$ is a normalizing constant. Using Markov chain Monte Carlo (MCMC) sampling, DAGs can be drawn from the posterior distribution, yielding a posterior sample $\mathcal{G}_1,\ldots,\mathcal{G}_R$.\\

In the BN literature, the marginal likelihood $p(\mathcal{D}|\mathcal{G})$ is often referred to as the score of the DAG $\mathcal{G}$. Two BN scoring metrics allow the marginal likelihood to be computed analytically:
\begin{itemize}
\item The {\em BDe scoring metric} \citep{HeckermanPrior,HeckermanPrior2,GeigerHeckGaussUAI2} is used to compute the marginal likelihood of DAGs over discrete (categorical) random variables. BDe stands for “Bayesian scoring metric for discrete networks having score equivalence.”
\item The {\em BGe scoring metric} \citep{HECKERMAN_ANNALS,HECKERMAN_ADDENDUM,GeigerHeckGaussUAI2,GeigerHeckGaussUAI} is used to compute the marginal likelihood of DAGs over Gaussian random variables. BGe stands for “Bayesian scoring metric for Gaussian networks having score equivalence.”
\end{itemize}
These two scoring metrics not only allow the marginal likelihood  to be computed analytically, but also ensure score equivalence, i.e. equivalent DAGs receive the same marginal likelihood value. This property is crucial for BNs, since equivalent graphs describe the same conditional dependencies and therefore should not be assigned different scores.

\subsection{Gaussian Bayesian networks with the BGe score}
\label{sec:gbn}
In Gaussian Bayesian networks (GBNs), the random vector ${\bf X}\in\mathbb{R}^n$ follows an $n$-dimensional Gaussian distribution with mean vector $\Bmu$ and covariance matrix $\Bsigma$. For a dataset $\mathcal{D}=\{{{\bf x}_1,\ldots,{\bf x}_m}\}$ we then have
\begin{equation}
    \label{EQ_VECTOR}
    p(\mathcal{D}| \Bmu,\Bsigma) \;=\; \prod_{i=1}^m \mathcal{N}_n({\bf x}_i | \Bmu,\Bsigma),
\end{equation}
where $\mathcal{N}_n({\bf x}|\Bmu,\Bsigma)$ denotes the density of the $n$-variate normal distribution $\mathcal{N}_n(\Bmu,\Bsigma)$ at ${\bf x}$.  A fully conjugate prior for the mean vector $\Bmu$ and the precision matrix ${\bf W} = \Bsigma^{-1}$ is given by the Normal–Wishart distribution:
\begin{eqnarray}
   \label{BGE_prior_old}
    {\bf W} & \sim & \mathcal{W}_n(\alpha_w,{\bf T}_{\dagger}), \\
    \nonumber
    \Bmu|{\bf W} & \sim & \mathcal{N}_n(\Bnu,(\alpha_{\mu} {\bf W})^{-1}),
\end{eqnarray}
where ${\bf T}_{\dagger}\in\mathbb{R}^{n\times n}$, $\Bnu\in\mathbb{R}^n$, $\alpha_w > n-1$, and $\alpha_{\mu}>0$  are hyperparameters. ${\bf T}_{\dagger}$ is the positive definite parametric matrix of the Wishart prior, $\Bnu$ is the mean vector of the Gaussian prior, and $\alpha_w$ and $\alpha_{\mu}$ can be interpreted as prior equivalent sample sizes \citep{HECKERMAN_ANNALS}. Since the prior is fully conjugate, the marginal likelihood 
\begin{equation}
\label{LIKELIHOOD_FULL}
    p(\mathcal{D}) = \int \int p(\mathcal{D}|\Bmu,{\bf W}^{-1})\cdot p(\Bmu|{\bf W}) \cdot p({\bf W})\; d\Bmu \;d{\bf W}
\end{equation}  can be computed analytically. However, the model specified in Eqs.~(\ref{EQ_VECTOR}--\ref{BGE_prior_old}) does not impose any conditional independence relations among the variables. That is, every pair of variables is dependent and remains dependent even when conditioning on any subset of the others. In terms of BNs, Eq.~(\ref{LIKELIHOOD_FULL}) therefore corresponds to the marginal likelihood of a complete DAG $\mathcal{G}_C$, symbolically $p(\mathcal{D}) = p(\mathcal{D}|\mathcal{G}_C)$. A complete DAG $\mathcal{G}_C$ has the maximal number of edges, implying that no pairwise independencies exist. For general DAGs, by contrast, the marginal likelihood incorporates the conditional independencies implied by the parent sets $\BPi_i$ through the factorization in Eq.~(\ref{EQ_FACTORIZATION}). \\
In their seminal works on GBNs \citep{HECKERMAN_ANNALS,HECKERMAN_ADDENDUM,GeigerHeckGaussUAI2,GeigerHeckGaussUAI}, Geiger and Heckerman extend the model in Eqs.~(\ref{EQ_VECTOR}--\ref{BGE_prior_old}) to non-complete DAGs. A DAG $\mathcal{G}$ specifies conditional independence relations among the variables, and it must be ensured that the precision matrix ${\bf W}$ is consistent with $\mathcal{G}$, i.e. that $\mathcal{G}$ and ${\bf W}$ imply the same conditional independencies. Geiger and Heckerman also show how to compute the marginal likelihood $p(\mathcal{D}|\mathcal{G})$ for any given DAG $\mathcal{G}$, and they refer to this marginal likelihood as the BGe score of $\mathcal{G}$.

\subsection{Mixture Gaussian Bayesian networks}
\label{sec:MIX}
In this section we assume that the data arise from a mixture of GBNs. In Section~\ref{sec:MIX_OLD} we consider the full-covariance mixture model, in which each component has its own Gaussian distribution with a separate covariance matrix. In Section~\ref{sec:MIX_NEW} we introduce the tied-covariance mixture model, in which all components share a common covariance matrix. The full-covariance model yields a marginal likelihood that factorizes into component-specific BGe scores. For the tied-covariance model we derive the marginal likelihood, which we refer to as the BGe score for tied-covariance mixture GBNs. \\
In both sections we let $z_i=k$ indicate that observation vector ${\bf x}_i$ is assigned to mixture component $k$. We assume there are $K$ components, and define the assignment vector ${\bf z}=(z_1,\dots,z_m)$, which allocates $m_k$ observations ${\bf x}_{k,l}$ ($l=1,\ldots,m_k$) to component $k$. Accordingly, we can write
$$\mathcal{D}= \{{\bf x}_{k,l}|k=1,\ldots,K;l=1,\ldots,m_k\},\; \mbox{where} \; \sum\limits_{k=1}^K m_k = m.$$
For each component $k$, we have the empirical mean vector and the scatter matrix (cross-product matrix):
\begin{equation}
\label{mean_scatter_defined}
\bar{{\bf x}}_{[k]}  :=  \frac{1}{m_k}\sum\limits_{l=1}^{m_k} {\bf x}_{k,l}, \; \mbox{and}\;\;
{\bf S}_{[k]}  :=  \sum\limits_{l=1}^{m_k}  ({\bf x}_{k,l} - \bar{{\bf x}}_{[k]}) ({\bf x}_{k,l} - \bar{{\bf x}}_{[k]})^{\transp}.  
\end{equation}

\subsubsection{Full-covariance mixture Gaussian Bayesian networks}
\label{sec:MIX_OLD}
Suppose each observation vector ${\bf x}_i$ ($i=1,\ldots,m$) arises from a Gaussian mixture distribution with $K$ components:  
\begin{equation}
    \label{x_mixture}
   p({\bf x}_i |\Btheta) \;=\; \sum_{k=1}^K \pi_k \, \mathcal{N}_n({\bf x}_i | \Bmu_k,\Bsigma_k),
\end{equation}
where $\Btheta := \{(\pi_1,\Bmu_1,\Bsigma_1),\ldots,(\pi_K,\Bmu_K,\Bsigma_K)\}$ is the set of model parameters, $\pi_k>0$ is the weight of component $k$, and
$\sum_{k=1}^K \pi_k = 1$.
On each pair $(\Bmu_k,{\bf W}_k)$, where ${\bf W}_k = \Bsigma_k^{-1}$, we impose the fully conjugate Normal--Wishart prior from Eq.~(\ref{BGE_prior_old}):  
\begin{eqnarray}
   \label{BGE_prior_mix_old}
   {\bf W}_k &\sim& \mathcal{W}_n(\alpha_{w,k},{\bf T}_{\dagger,k}), \\
   \nonumber
   \Bmu_k | {\bf W}_k &\sim& \mathcal{N}_n\!\left(\Bnu_k,(\alpha_{\mu,k} {\bf W}_k)^{-1}\right),
\end{eqnarray}
where ${\bf T}_{\dagger,k}\in\mathbb{R}^{n\times n}$ is positive definite, $\Bnu_k \in \mathbb{R}^n$, $\alpha_{w,k} > n-1$, and $\alpha_{\mu,k} > 0$ are hyperparameters. As before, $\alpha_{w,k}$ and $\alpha_{\mu,k}$ can be interpreted as prior equivalent sample sizes \citep{HECKERMAN_ANNALS}. \\
For a given assignment vector ${\bf z}$ and a complete DAG $\mathcal{G}_C$, the component-specific posterior distributions are
\begin{eqnarray}
\label{POSTERIOR_OLD}
{\bf W}_k|(\mathcal{D},{\bf z},\mathcal{G}_C) & \sim &  \mathcal{W}_n( \alpha_{w,k}+m_k, {\bf T}_{\ddagger,k} ),  \\
\nonumber
\Bmu_k|(\mathcal{D},{\bf z},\mathcal{G}_C,{\bf W}_k) & \sim &  \mathcal{N}_n\left( \Bmu_k^{\diamond}, \left((\alpha_{\mu,k}+m_k){\bf W}_k\right)^{-1}\right),  
\end{eqnarray}
with
\begin{eqnarray} 
\label{T_ddagger_old}
{\bf T}_{\ddagger,k}  & := & {\bf T}_{\dagger,k} +  {\bf S}_{[k]} + \frac{\alpha_{\mu,k}\cdot m_k}{\alpha_{\mu,k}+m_k} (\Bnu_k - \bar{{\bf x}}_{[k]}) (\Bnu_k - \bar{{\bf x}}_{[k]} )^{\transp}, \\
\nonumber
\Bmu_k^{\diamond} & := & \frac{\alpha_{\mu,k} \Bnu_k + m_k \bar{{\bf x}}_{[k]}  }{\alpha_{\mu,k}+m_k}.
\end{eqnarray}
 For the marginal likelihood we obtain:
{\footnotesize
\begin{equation}
 \label{MARGINAL_1}
p(\mathcal{D}|{\bf z},\mathcal{G}_C)  =  \prod\limits_{k=1}^K \left( \pi^{-\frac{n\cdot m_k}{2}} \cdot
\left(  \frac{\alpha_{\mu,k}}{\alpha_{\mu,k}+m_k}   \right)^{\frac{n}{2}}
   \cdot \frac{\Gamma_n(\frac{\alpha_{w,k}+m_k}{2})}{\Gamma_n(\frac{\alpha_{w,k}}{2})} \cdot   
  \frac{  \det({\bf T}_{\dagger,k})^{\alpha_{w,k}/2}}{  \det({\bf T}_{\ddagger,k})^{(\alpha_{w,k}+m_k)/2}}\right),
\end{equation}
}
where $\Gamma_n(\cdot)$ denotes the $n$-dimensional multivariate gamma function. From Eq.~(\ref{MARGINAL_1}) it can be seen that the marginal likelihood factorizes into $K$ terms. For any $l$-dimensional variable subset $L \subset \{X_1,\ldots,X_n\}$ we have
{\footnotesize
\begin{eqnarray}
 \nonumber
p(\mathcal{D}^L|{\bf z},\mathcal{G}^L_C)  =  \prod\limits_{k=1}^K \left( \pi^{-\frac{l\cdot m_k}{2}} \cdot
\left(  \frac{\alpha_{\mu,k}}{\alpha_{\mu,k}+m_k}   \right)^{\frac{l}{2}}
   \cdot \frac{\Gamma_l(\frac{\alpha_{w,k}-n+l+m_k}{2})}{\Gamma_l(\frac{\alpha_{w,k}-n+l}{2})} \cdot   
  \frac{  \det({\bf T}_{\dagger,k}^{L,L})^{\alpha_{w,k}/2}}{  \det({\bf T}_{\ddagger,k}^{L,L})^{(\alpha_{w,k}+m_k)/2}}\right), \\
   \label{EQ_MARGINAL2_OLD}
\end{eqnarray}
}
where $\mathcal{D}^L$ denotes the data restricted to the $l$ variables in $L$, $\mathcal{G}_C^L$ is the complete DAG on those $l$ nodes, and ${\bf T}_{\dagger,k}^{L,L}$ and ${\bf T}_{\ddagger,k}^{L,L}$ are the submatrices of ${\bf T}_{\dagger,k}$ and ${\bf T}_{\ddagger,k}$ corresponding to the variables in $L$.

\subsubsection{Tied-covariance mixture Gaussian Bayesian networks}
\label{sec:MIX_NEW}
In the tied-covariance mixture model, in contrast to the full-covariance case in Eq.~(\ref{x_mixture}), we assume that all components share the same covariance matrix $\Bsigma$, while only the mean vectors $\Bmu_k$ are component-specific:
\begin{equation}
    \label{x_mixture_new}
   p({{\bf x}}_i|\Btheta)     =  \sum_{k=1}^K \pi_k \; \mathcal{N}_n({\bf x}_i|\Bmu_k,\Bsigma). 
\end{equation}
We impose a Wishart prior on the precision matrix ${\bf W}=\Bsigma^{-1}$.  
Conditional on ${\bf W}$, we then assign independent Gaussian priors to the component-specific mean vectors $\Bmu_k$ ($k=1,\ldots,K$):  
\begin{equation}
\label{mixBGe_PRIOR_NEW}
   {\bf W} \sim \mathcal{W}_n(\alpha_w,{\bf T}_{\dagger}), 
   \qquad
   \Bmu_k | {\bf W} \sim \mathcal{N}_n\!\left(\Bnu_k,(\alpha_{\mu,k}{\bf W})^{-1}\right),
\end{equation}
where ${\bf T}_{\dagger}\in\mathbb{R}^{n\times n}$ is positive definite, $\Bnu_k \in \mathbb{R}^n$, $\alpha_w > n-1$, and $\alpha_{\mu,k} > 0$ are hyperparameters.  
As before, $\alpha_w$ and $\alpha_{\mu,k}$ can be interpreted as prior equivalent sample sizes \citep{HECKERMAN_ANNALS}.\\

The shared precision matrix ${\bf W}$ links the mixture components, so that we no longer have $K$ independent component-specific Gaussian models. Nevertheless, the posterior distribution and the marginal likelihood can still be computed analytically. Since the derivations are lengthy and technical, they are presented in {\bf Part~A of the supplementary material}; here we report only the results.
For a complete DAG $\mathcal{G}_C$ and a given assignment vector ${\bf z}$, the tied-covariance model yields the following posterior distribution: 
\begin{eqnarray}
\nonumber
{\bf W}|(\mathcal{D},{\bf z},\mathcal{G}_C) & \sim &  \mathcal{W}_n( \alpha_{w}+m, {\bf T}_{\ddagger} ),  \\
\label{POSTERIOR_NEW}
\Bmu_k|(\mathcal{D},{\bf z},\mathcal{G}_C,{\bf W}) & \sim &  \mathcal{N}_n\left( \Bmu_k^{\diamond}, \left((\alpha_{\mu,k}+m_k){\bf W}\right)^{-1}\right),  
\end{eqnarray}
with
\begin{equation} 
\label{T_ddagger}
{\bf T}_{\ddagger}   :=  {\bf T}_{\dagger} + \sum\limits_{k=1}^K {\bf T}_{[k]},\; \mbox{and}\;
\Bmu_k^{\diamond}  :=  \frac{\alpha_{\mu,k} \Bnu_k + m_k \bar{{\bf x}}_{[k]}  }{\alpha_{\mu,k}+m_k},
\end{equation}
$$\mbox{where}\; {\bf T}_{[k]} :=  {\bf S}_{[k]} + \frac{\alpha_{\mu,k}\cdot m_k}{\alpha_{\mu,k}+m_k} (\Bnu_k - \bar{{\bf x}}_{[k]}) (\Bnu_k - \bar{{\bf x}}_{[k]} )^{\transp}.$$

\noindent  For a complete DAG $\mathcal{G}_C$, the marginal likelihood is 
\begin{equation}
 \label{MARGINAL_2}
p(\mathcal{D}|{\bf z},\mathcal{G}_C)  =   \pi^{-\frac{n\cdot m}{2}} \cdot
\prod\limits_{k=1}^K \left(  \frac{\alpha_{\mu,k}}{\alpha_{\mu,k}+m_k}   \right)^{\frac{n}{2}}
   \cdot \frac{\Gamma_n(\frac{\alpha_w+m}{2})}{\Gamma_n(\frac{\alpha_w}{2})} \cdot   
  \frac{  \det({\bf T}_{\dagger})^{\alpha_w/2}}{  \det({\bf T}_{\ddagger})^{(\alpha_w+m)/2}}.
\end{equation}

\noindent For an $l$-dimensional subset $L\subset\{X_1,\ldots,X_n\}$, this implies 
{\small
\begin{eqnarray}
  \nonumber
p(\mathcal{D}^L|{\bf z},\mathcal{G}_C^L) & = &  \pi^{-\frac{l\cdot m}{2}} \cdot
\prod\limits_{k=1}^K \left(  \frac{\alpha_{\mu,k}}{\alpha_{\mu,k}+m_k}   \right)^{\frac{l}{2}}
   \cdot \frac{\Gamma_l(\frac{\alpha_w-n+l+m}{2})}{\Gamma_l(\frac{\alpha_w-n+l}{2})} \cdot   
  \frac{  \det\left({\bf T}_{\dagger}^{L,L}\right)^{\alpha_w/2}}{  \det\left({\bf T}_{\ddagger}^{L,L}\right)^{(\alpha_w+m)/2}}, \\
  \label{EQ_MARGINAL2}
\end{eqnarray}
}
 where $\mathcal{D}^L$ is the data restricted to the variables in $L$, $\mathcal{G}_C^L$ is the complete DAG on $L$, and ${\bf T}_{\dagger}^{L,L}$ and ${\bf T}_{\ddagger}^{L,L}$ are the corresponding submatrices of ${\bf T}_{\dagger}$ and ${\bf T}_{\ddagger}$.

\subsubsection{Comparison of full-covariance and tied-covariance models}
\label{sec:MIX_COMPARISON}
To highlight the differences between the two marginal likelihoods, we set ${\bf T}_{\dagger,k} = {\bf T}_{\dagger}$, $\alpha_{w,k} = \alpha_w$, and $\alpha_{\mu,k} = \alpha_{\mu}$. For complete DAGs $\mathcal{G}_C$, the marginal likelihoods for the full-covariance (M1) and tied-covariance (M2) mixture GBNs are then given by
{\small
\begin{eqnarray}
\nonumber
p(\mathcal{D}|{\bf z},\mathcal{G}_C,\mathcal{M}_1) & = & \tilde{c} \; \cdot \; \prod\limits_{k=1}^K \frac{\Gamma_n(\frac{\alpha_{w}+m_k}{2})}{\Gamma_n(\frac{\alpha_{w}}{2})}  \;\; \cdot   
\;\;\;\; \prod\limits_{k=1}^K  \frac{  \det({\bf T}_{\dagger})^{\alpha_{w}/2}}{  \det({\bf T}_{\dagger} + {\bf T}_{[k]})^{(\alpha_{w}+m_k)/2}}, \\
 \nonumber
 & & \\
 \nonumber
p(\mathcal{D}|{\bf z},\mathcal{G}_C,\mathcal{M}_2) & = &  \tilde{c}\;
   \cdot \;\frac{\Gamma_n(\frac{\alpha_w+\sum_k m_k}{2})}{\Gamma_n(\frac{\alpha_w}{2})} \;\;\;\; \cdot   
  \frac{  \det({\bf T}_{\dagger})^{\alpha_w/2}}{  \det({\bf T}_{\dagger} + \sum_k {\bf T}_{[k]})^{(\alpha_w+\sum_k m_k)/2}},
\end{eqnarray}}
where {\footnotesize $\tilde{c} := \pi^{-\frac{n\cdot m}{2}} \cdot
\prod\limits_{k=1}^K  \left( \frac{\alpha_{\mu}}{\alpha_{\mu}+m_k}   \right)^{\frac{n}{2}}$} is a constant. For $K=1$, both marginal likelihoods reduce to the standard marginal likelihood of a Gaussian model with a Normal–Wishart prior, i.e. the BGe score of a homogeneous GBN.

\subsubsection{Marginal likelihood and posterior distribution of any DAG}
\label{SEC:ANY_DAG}
For both mixture models, the marginal likelihood for any DAG $\mathcal{G}$ can be computed using the factorization of  \cite{GeigerHeckGaussUAI, HECKERMAN_ANNALS, 
 HECKERMAN_ADDENDUM}:
\begin{equation}
   \label{EQ_MARGINAL_ANY}
   p(\mathcal{D}|{\bf z},\mathcal{G}) =  \prod\limits_{i=1}^n \frac{ p(\mathcal{D}^{\{X_i,\BPi_i\}}|{\bf z},\mathcal{G}_C^{\{X_i,\BPi_i\}})} {p(\mathcal{D}^{\{\BPi_i\}}|{\bf z},\mathcal{G}_C^{\{\BPi_i\}})},
 \end{equation}  
    where $\BPi_i$ denotes the parent node set of $X_i$ implied by $\mathcal{G}$, and $p(\mathcal{D}^{\{X_i,\BPi_i\}}|{\bf z},\mathcal{G}_C^{\{X_i,\BPi_i\}})$  and $p(\mathcal{D}^{\{\BPi_i\}}|{\bf z},\mathcal{G}_C^{\{\BPi_i\}})$ are marginal likelihoods for the subsets $L_{i,1}:=\{X_i,\BPi_i\}$ and $ L_{i,2}:= \{\BPi_i\}$. These marginal likelihoods for variable subsets  can be computed with either Eq.~(\ref{EQ_MARGINAL2_OLD}) or Eq.~(\ref{EQ_MARGINAL2}), depending on whether the full-covariance or the tied-covariance approach is applied. \\
Covariance matrices $\Bsigma_k = {\bf W}_k^{-1}$ sampled from Eq.~(\ref{POSTERIOR_OLD}) or $\Bsigma = {\bf W}^{-1}$ sampled from Eq.~(\ref{POSTERIOR_NEW}) correspond to complete DAGs $\mathcal{G}_C$, since they do not encode any conditional independence relations. To obtain posterior samples of covariance matrices that are consistent with the conditional independencies implied by an arbitrary DAG $\mathcal{G}$, we apply the algorithm of \cite{Gryze2023}.

\subsection{Joint posterior inference of class assignments and DAGs}
\label{sec:mixture}
In Section~\ref{sec:MIX} we assumed that the number of mixture components $K$ and the assignment vector ${\bf z}$ were fixed and known. To infer both from the data, we place a Dirichlet prior with parameters $\alpha_1=\cdots=\alpha_K=1$ on the mixture weights $\pi_1,\ldots,\pi_K$, and, following  \cite{newman2025fast}, we exclude empty components.\footnote{That is, we exclude all assignment vectors ${\bf z}$ for which $m_k=0$ for some $k\in\{1,\ldots,K\}$.} 
For any assignment vector ${\bf z}=(z_1,\ldots,z_m)$ with component counts $(m_1,\ldots,m_K)$ such that $m_k>0$ for all $k=1,\ldots,K$, the induced distribution on assignments (obtained by integrating out the mixture weights under a Dirichlet$(1,\ldots,1)$ prior) is the Dirichlet–multinomial distribution conditioned on non-empty components \citep{newman2025fast}:
\begin{equation}
p({\bf z}) = \binom{m-1}{K-1}^{-1} \cdot \frac{\prod_{k=1}^K m_k!}{m!}.
\end{equation}
Conditional on ${\bf z}$, the posterior distribution of the mixture weights is
\begin{equation}
p(\pi_1,\ldots,\pi_K|{\bf z}) = \frac{\Gamma(m+K)}{\prod_{k=1}^K \Gamma(m_k+1)}  \cdot \prod_{k=1}^K \pi_k^{m_k},
\end{equation}
which is the Dirichlet density with parameters $\tilde{\alpha}_k = m_k+1$ ($k=1,\ldots,K$). \\
Furthermore, since empty components are ruled out, the number of mixture components $K$ is fully determined by the assignment vector ${\bf z}$. To encourage sparser models, we place a Poisson prior with rate parameter $\lambda>0$ on $K$
$$ p(K) = \frac{\lambda^K \cdot e^{-\lambda}}{K!}.$$

For the joint posterior distribution of a DAG $\mathcal{G}$ and the assignment vector ${\bf z}$, we have:
\begin{equation}
\label{EQ_POSTERIOR}
    p(\mathcal{G},{\bf z}|\mathcal{D}) \propto p(\mathcal{D}|\mathcal{G},{\bf z}) \cdot p(\mathcal{G}) \cdot p({\bf z}) \cdot p(K),
\end{equation} 
where $p(\mathcal{D}|\mathcal{G},{\bf z})$ is the marginal likelihood from Eq.~(\ref{EQ_MARGINAL_ANY}),  
$p(\mathcal{G})$ is the prior on DAGs, $p({\bf z})$ is the Dirichlet–multinomial distribution over assignments, 
and the number of mixture components $K$ is determined by ${\bf z}$. 

\subsection{MCMC sampling scheme for mixture GBNs}
\label{sec:mcmc}
To generate samples from the joint posterior distribution in Eq.~\eqref{EQ_POSTERIOR}, 
we combine the classical structure MCMC sampler for Bayesian networks 
\citep{MadiganYork,GiudiciMCMC} with the recently proposed fast collapsed Gibbs sampler 
for mixture models by \cite{newman2025fast}.\footnote{The fast Gibbs sampler of 
\cite{newman2025fast} applies to mixture models with a categorical–Dirichlet prior and 
analytically computable marginal likelihoods.} 
Given the current state $[\mathcal{G},{\bf z}]$, where $\mathcal{G}$ is a DAG on the $n$ variables 
and ${\bf z}=(z_1,\ldots,z_m)$ is the assignment vector of the mixture model, 
our MCMC scheme alternates between two sampling steps:\\

\noindent {\bf Step 1: Metropolis-Hastings move on the DAG}: \\
Given the current state $[\mathcal{G},{\bf z}]$, let $N(\mathcal{G})$ denote the neighborhood of the  DAG $\mathcal{G}$, i.e. the set of all DAGs that can be reached from $\mathcal{G}$ by adding, deleting or reversing one single edge. We propose a candidate $\mathcal{G}^{\star}\in N(\mathcal{G})$ by selecting a neighbor at random.  The acceptance probability of the move is:
\begin{equation}
  \label{EQ_A_1}
A([\mathcal{G},{\bf z}]\rightarrow [\mathcal{G}^{\star},{\bf z}]) = \min\left\{1,\; \frac{p(\mathcal{D}|\mathcal{G}^{\star},{\bf z})}{p(\mathcal{D}|\mathcal{G},{\bf z})} \cdot \frac{p(\mathcal{G}^{\star})}{p(\mathcal{G})} \cdot \mbox{HR}(\mathcal{G},\mathcal{G}^{\star}) \right\} 
\end{equation}
    with the Hastings ratio $\mbox{HR}({\mathcal{G}},\mathcal{G}^{\star})=\frac{|N(\mathcal{G})|}{|N(\mathcal{G}^{\star})|}$, where $|\cdot|$ denotes the cardinality of the corresponding neighborhood.  
      If the move is accepted, the state is updated to $[\mathcal{G}^{\star},{\bf z}]$; otherwise it remains at $[\mathcal{G},{\bf z}]$.\\

\noindent {\bf Step 2: Collapsed Gibbs move on the assignment vector}: \\
Given the current state $[\mathcal{G},\mathbf z]$, the assignment vector $\mathbf z=(z_1,\ldots,z_m)$ 
implies $K$ mixture components. Let $\mathbf z_{-i}$ denote the assignment vector with the $i$-th 
observation removed, i.e.\ $\mathbf z_{-i}=(z_1,\ldots,z_{i-1},z_{i+1},\ldots,z_m)$.
\begin{itemize}
\item First, choose one of the $K$ components uniformly at random. From this component $k$, 
select one of its $m_k$ observations uniformly at random. This yields an observation index 
$i \in \{1,\ldots,m\}$ assigned to component $k$. Remove observation $i$ from $k$. 
If $m_k=1$, the removal leaves an empty component; in this case, delete component $k$ 
and decrement all subsequent component indices $k+1,\ldots,K$ by one.
\item After removal (and possible deletion), let $\tilde{K}$ be the current number of components.
For $s=1,\ldots,\tilde{K}+1$, compute the unnormalized weights:
\[
w_s =
\begin{cases}  
\displaystyle \frac{m-\tilde{K}}{\tilde{K}} \cdot p(\tilde{K}) \cdot p(\mathcal{D}| \mathcal{G},\mathbf z_{-i}, z_i = s), & s=1,\ldots,\tilde{K}, \\[0.8em]
\displaystyle \tilde{K} \cdot p(\tilde{K}+1) \cdot p(\mathcal{D}| \mathcal{G},\mathbf z_{-i}, z_i = \tilde{K}+1), & s=\tilde{K}+1 .
\end{cases}
\]
and the corresponding probabilities
\[
p_s = \frac{w_s}{\sum_{\nu=1}^{\tilde{K}+1} w_\nu}, \qquad s=1,\ldots,\tilde{K}+1 .
\]
\item Finally, re-allocate observation $i$ to one of the components $1,\ldots,\tilde{K}+1$, choosing 
component $s$ with probability $p_s$. The new state is then $[\mathcal{G},\mathbf z^{\star}]$, where 
$\mathbf z^{\star} = (z_1,\ldots,z_{i-1},s,z_{i+1},\ldots,z_m)$.
\end{itemize}

Repeatedly performing the two sampling steps yields a Markov chain whose stationary 
distribution is the joint posterior in Eq.~(\ref{EQ_POSTERIOR}). 
We run the algorithm for $T$ iterations. To account for burn-in and to reduce 
autocorrelation, we discard the first 50\% of the output and thin the remaining 
50\% by a factor of $\xi$. This results in a posterior sample of size 
$R = \left\lfloor \tfrac{T}{2\xi} \right\rfloor$:
\begin{equation}
\label{eq:post_sample}
[\mathcal{G}^{(1)},{\bf z}^{(1)}], \ldots, [\mathcal{G}^{(R)},{\bf z}^{(R)}].
\end{equation}

\subsection{Evaluation of the posterior sample}
\label{sec:evaluation}
We use the posterior sample from Eq.~(\ref{eq:post_sample}) for a range of purposes: 
\begin{itemize}
\item From the posterior samples of the assignment vectors ${\bf z}^{(1)},\ldots,{\bf z}^{(R)}$
we compute the {\bf pairwise co-allocation probabilities}, defined as the probability
that two observations $i$ and $j$ are assigned to the same mixture component.
These probabilities can be visualized in a heatmap, where block structures indicate groups of observations that are consistently co-assigned across the posterior sample.
\item From the sampled DAGs, we compute the {\bf marginal edge posterior probability} of an edge 
$X_i \rightarrow X_j$ as
\[
\hat{p}_{i,j}  :=  \frac{1}{R} \sum_{r=1}^R 
   \mathcal{I}_{X_i\rightarrow X_j}(\mathcal{G}^{(r)}),
\]
where $\mathcal{I}_{X_i\rightarrow X_j}(\mathcal{G})=1$ if the CPDAG of $\mathcal{G}$ contains 
either the directed edge $X_i\rightarrow X_j$ or the undirected edge $X_i - X_j$, and 
$\mathcal{I}_{X_i\rightarrow X_j}(\mathcal{G})=0$ otherwise.\footnote{Because DAGs fall into 
equivalence classes, each DAG is first replaced by its CPDAG. If the CPDAG has the undirected 
edge $X_i - X_j$, we interpret it as a bidirectional edge $X_i \leftrightarrow X_j$. Under 
this interpretation, the undirected edge implies that $X_i \rightarrow X_j$ is present.}
\item Imposing a threshold $\psi\in[0,1]$ on the edge probabilities $\hat{p}_{i,j}$ yields the {\bf network prediction} 
$\hat{\mathcal{G}}_{\psi}$, which contains only those edges with posterior probability 
exceeding~$\psi$.
\item To assess the {\bf network reconstruction accuracy} beyond a fixed threshold $\psi$, we evaluate the {\bf area under the precision–recall curve (AUC)}. Given the true CPDAG, each edge posterior probability $\hat{p}_{i,j}$ serves as a score for predicting whether the edge is present. Undirected edges in the true CPDAG are treated as undirected, meaning that either orientation $X_i\rightarrow X_j$ or $X_j\rightarrow X_i$ is counted as a correct recovery; this interpretation is consistent with the edge score computation. Varying the threshold $\psi$ from $1$ down to $0$ yields a sequence of precision–recall pairs, where precision is the proportion of predicted edges that are true, and recall is the proportion of true edges that are recovered. The AUC$\in[0,1]$ summarizes this curve into a single measure, with higher values indicating better agreement between the inferred edge probabilities and the true CPDAG. 
\item For every posterior sample $[\mathcal{G}^{(r)},{{\bf z}}^{(r)}]$ 
we can draw model parameters with covariance matrices coherent with $\mathcal{G}^{(r)}$ 
\citep{Gryze2023}:
\[
\Btheta^{(r)} = 
\begin{cases} 
 \{(\pi_1^{(r)},\Bmu_{1}^{(r)},\Bsigma_{1}^{(r)}),\ldots,(\pi_{K_r}^{(r)},\Bmu_{K_r}^{(r)},\Bsigma_{K_r}^{(r)}) \}, 
   & \text{(full-covariance)}, \\[1em]
 \{(\pi_1^{(r)},\Bmu_{1}^{(r)}),\ldots,(\pi_{K_r}^{(r)},\Bmu_{K_r}^{(r)})\},\;\Bsigma^{(r)}, 
   & \text{(tied-covariance)},
\end{cases}
\]
where the number of components $K_r$ is determined by ${\bf z}^{(r)}$.
\item For a new dataset with $m^{\star}$ observations
$\mathcal{D}^{\star} := \{{\bf x}_1^{\star},\ldots,{\bf x}_{m^{\star}}^{\star}\}$, 
we compute the {\bf geometric mean predictive probability} using Monte Carlo integration. 
The likelihood of the new data is averaged over the posterior samples of the parameters 
obtained from $\mathcal{D}$ and then normalized by the $m^{\star}$-th root:
\[
p(\mathcal{D}^{\star}\mid\mathcal{D})  
:= \left[ \frac{1}{R} \sum_{r=1}^R   
   \prod_{i=1}^{m^{\star}}   p({\bf x}_i^{\star}|\Btheta^{(r)}) \right]^{1/m^{\star}},
\]
where
\[
p({\bf x}^{\star}_i|\Btheta^{(r)}) =
\begin{cases} 
 \sum_{k=1}^{K_r}  \pi_k^{(r)}\, \mathcal{N}_n({\bf x}^{\star}_i|\Bmu_k^{(r)},\Bsigma_k^{(r)}), 
   & \text{(full-covariance)}, \\[0.6em]
 \sum_{k=1}^{K_r}  \pi_k^{(r)}\, \mathcal{N}_n({\bf x}^{\star}_i|\Bmu_k^{(r)},\Bsigma^{(r)}), 
   & \text{(tied-covariance)}.
\end{cases}
\]
\end{itemize}

  \subsection{Implementation details}
  \label{sec:implementation}
If no prior knowledge about the network is available, we assume all DAGs $\mathcal{G}$ 
to be equally likely. The hyperparameters of the Normal–Wishart priors are set to yield 
uninformative (weak) priors. Specifically, we take 
${\bf T}_{\dagger,k}={\bf T}_{\dagger}={\bf I}$ and $\Bnu_k = \Bnu = {\bf 0}$, 
where ${\bf I}$ is the identity matrix and ${\bf 0}$ the zero vector. 
For the equivalent sample size parameters we choose very small values: 
$\alpha_{w,k}=\alpha_{w}=n+1$ (recall that $\alpha_w > n-1$ is required) and 
$\alpha_{\mu,k}=\alpha_{\mu}=1$. 
This makes the Normal–Wishart priors weakly informative. For example, in the tied-covariance model the priors are:
${\bf W} \sim \mathcal{W}_n(n+1,{\bf I})$ and for each component $k$: 
$\Bmu_k|{\bf W} \sim \mathcal{N}_n({\bf 0},{\bf W}^{-1})$. Finally, we set the rate parameter of the Poisson prior on the number of mixture components 
to $\lambda=1$. \\
To assess convergence, we selected a few representative datasets and ran independent 
MCMC simulations on each of them. Standard convergence diagnostics were applied, 
such as trace plots and scatter plots of the edge scores \citep{Gryze2023,Gryze2024}, 
to compare the outputs of the independent runs. We determined the required number of 
iterations $T$ as the point at which the trace plots reached stable plateaus and the 
independent runs produced nearly identical marginal posterior probabilities. \\

The results reported in Sections~\ref{sec:results1} and \ref{sec:results2} are based on 
$250{,}000$ iterations, while those in Section~\ref{sec:results3} are based on 
$1{,}000{,}000$ iterations. In each case, the first $T/2$ iterations were discarded 
as burn-in, and a thinning factor $\xi$ was applied so that the final posterior sample 
consisted of $R=500$ draws from the second half of the Markov chain. \\

{\bf Software:} Upon acceptance, we will share the {\em R} code at \\ \url{https://github.com/MarcoAndreas/MIX-GBN}

\section{Data}
\label{sec:data}
We compare model performance on both simulated and real-world datasets. 
For the simulated data, the true networks (CPDAGs) are known, allowing us to 
evaluate network reconstruction accuracy using AUC and rSHD scores. 
For the real-world datasets, where the true network structures are unknown, 
we instead assess performance through predictive probabilities.

\subsection{Simulated network data}
\label{sec:data1}
We assume that the network domain consists of $n$ variables 
$X_1,\ldots,X_n$ arranged in topological order. 
This yields $\zeta = n(n-1)/2$ possible directed edges $X_j \rightarrow X_i$ with $j<i$. 
Each edge is assumed to be present independently with probability $p$, 
and we set $p = 20/\zeta$ so that the expected number of edges equals~20. In a Gaussian Bayesian network with zero mean vector, $\Bmu = {\bf 0}$, 
each observation $v$ satisfies
\begin{equation}
\label{eq:data}
x_{i,v} = \sum_{j \in \BPi_i} b_{i,j} x_{j,v} + \epsilon_{i,v},
\end{equation}
where $x_{i,v}$ is the value of variable $X_i$, 
$\BPi_i \subset \{X_1,\ldots,X_{i-1}\}$ is its parent set, 
and $\epsilon_{i,v} \sim \mathcal{N}(0,\sigma_i^2)$ denotes Gaussian noise. Eq.~(\ref{eq:data}) involves the parameters $\{b_{i,j}: j \in \BPi_i\}$ and $\sigma_i$. 
We sample these parameters from uniform distributions on $[1,2]$ and assign a random sign 
to each $b_{i,j}$. Finally, we rescale the parameters so that their Euclidean norm equals~1:
$$\sigma_i^2 + \sum_{j \in \BPi_i} b_{i,j}^2 = 1.$$
The joint distribution of ${\bf X}=(X_1,\ldots,X_n)^{\transp}$ is Gaussian with covariance 
matrix $\Bsigma$ determined by the rescaled parameters $\{b_{i,j}: j \in \BPi_i\}$ 
and $\sigma_i$ ($i=1,\ldots,n$) \citep{SHACHTER}. 
We generate $m$ observations
$$ \tilde{{\bf x}}_1,\ldots,\tilde{{\bf x}}_m \sim \mathcal{N}_n({\bf 0},\Bsigma).$$

We assume $K$ mixture components, each associated with a mean vector $\Bmu_k$. 
The mean vectors are sampled from standard Gaussian distributions:
$$\Bmu_1,\ldots,\Bmu_K \sim \mathcal{N}_n({\bf 0},{\bf I}).$$

Each observation $\tilde{{\bf x}}_v$ is then randomly assigned to a component $k$, 
with equal probabilities $\pi_k=1/K$.  If $\tilde{{\bf x}}_v$ is assigned to  $k$, we add the corresponding mean vector, 
so that ${\bf x}_v \leftarrow \tilde{{\bf x}}_v + \Bmu_k$.
Thus, each ${\bf x}_v$ is drawn from the tied-covariance mixture GBN in Eq.~(\ref{x_mixture_new}), 
with equal mixture weights.

\subsection{Benchmark datasets}
\label{sec:data2}

We use four benchmark datasets that are widely used in the statistics and machine 
learning literature. Within each dataset, all $n$ continuous variables are 
$z$-score standardized to have mean~0 and variance~1.\footnote{The data can, for instance, be obtained from 
\url{https://www.kaggle.com/datasets}.}

\subsubsection{Fish}
The Fish Market dataset \citep{FishMarketKaggle} contains $m=159$ fish samples 
from a market in Finland, belonging to $K=7$ species: 
{\em Bream}, {\em Roach}, {\em Whitefish}, {\em Parkki}, {\em Perch}, {\em Pike}, and {\em Smelt}. 
The dataset has $n=6$ variables: weight, vertical length, diagonal length, 
cross length, height, and diagonal width. 
Although often presented as real-world data, it remains unclear whether 
these fish were actually caught in the wild.

\subsubsection{Iris}
The classic Iris dataset \citep{IRIS} contains $m_i=50$ samples from each of 
$K=3$ species: {\em I. setosa}, {\em I. versicolor}, and {\em I. virginica}. 
It has $n=4$ variables: sepal length, sepal width, petal length, and petal width. 

\subsubsection{Wine}
The Wine dataset \citep{Forina1986} contains chemical analyses of $m=178$ wines 
from $K=3$ cultivars of the same Italian region. 
It includes $n=13$ variables: alcohol, malic acid, ash, alkalinity of ash, magnesium, 
total phenols, flavonoids, nonflavonoid phenols, proanthocyanins, color intensity, 
hue, OD280/OD315 (of diluted wines), and proline.

\subsubsection{Seeds}
The Wheat Seeds Kernel dataset \citep{Charytanowicz2010} contains $m_i=70$ kernel 
samples from each of $K=3$ wheat varieties: {\em Kama}, {\em Rosa}, and {\em Canadian}. 
It has $n=7$ variables: area, perimeter, compactness, kernel length, kernel width, asymmetry coefficient,  and kernel groove length.

\section{Results}
\label{sec:results}
In Section~\ref{sec:results1} we use simulated data to compare model performance 
in terms of network reconstruction accuracy.  The simulation study consists of two parts: 
first, we assume that the class labels are known;  second, we treat the class labels as latent and infer them from the data. 
In Section~\ref{sec:results2} we evaluate the models on the four benchmark datasets 
introduced in Section~\ref{sec:data2}. 
Since the true networks are unknown, performance is assessed using predictive probabilities. 
In Section~\ref{sec:results3} we evaluate the four benchmark datasets 
with the new tied-covariance mixture GBN model (M2) and present the empirical results.

\subsection{Simulated data}
\label{sec:results1}
We generate data as described in Section~\ref{sec:data1}. 
We cross four sample sizes, $m \in \{100,200,400,800\}$, with four numbers of mixture 
components, $K \in \{2,4,8,16\}$. 
For each $(m,K)$ combination we generate 25 independent replicates, 
resulting in a total of 400 simulated datasets.\\

\noindent {\bf Part 1:} First, we assume that the class labels are known, 
i.e. the $m$ observations are correctly allocated to the $K$ mixture components. 
In this setting, we compare the performance of five models.
\begin{itemize}
\item {\bf H}: The homogeneous Gaussian Bayesian network (GBN), using the BGe score. 
All $m$ observations are merged into a single component.  

\item {\bf M1}: The {\em full-covariance} mixture model, using the product of 
component-specific BGe scores from Section~\ref{sec:MIX_OLD}.  

\item {\bf M2}: The {\em tied-covariance} mixture model, using the newly derived 
modified BGe score from Section~\ref{sec:MIX_NEW}.  

\item {\bf clG}: The conditional linear Gaussian model for hybrid data, using the 
implementation in the {\bf R} package {\em bnlearn} \citep{SCUTARI_BOOK,SCUTARI_NEERLANDICA,SCUTARI_REASONING}.  

\item {\bf mBGe}: The mean-adjusted Gaussian Bayesian network for hybrid data, using 
the approach from \cite{Gryze2025}.
\end{itemize}

\begin{figure}[p]
\vspace{3mm}
 \begin{center}
\includegraphics[width=0.98\linewidth]{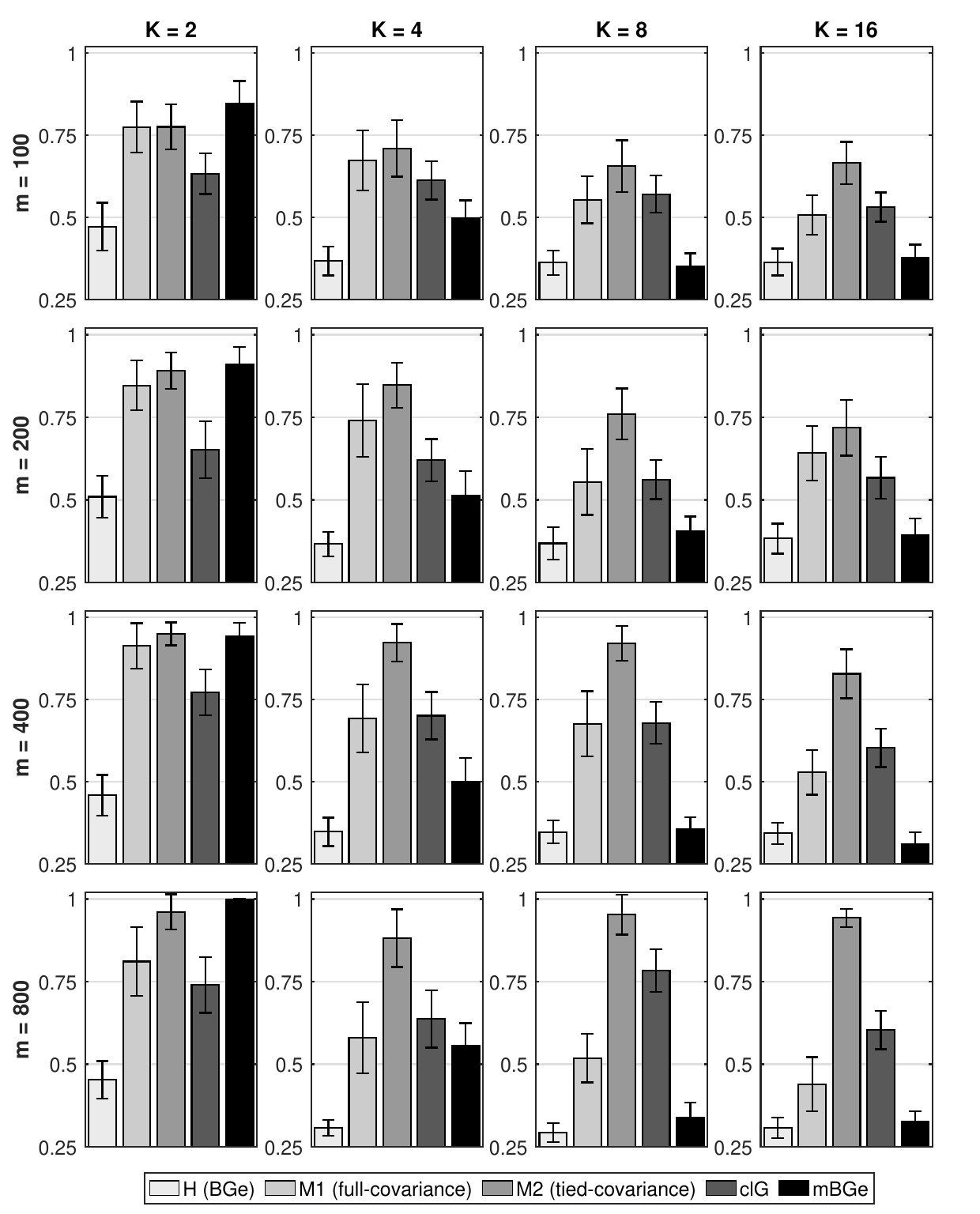} 
  \vspace{-0.35cm}
\caption {{\bf Areas under the Precision-Recall Curve (AUCs) for Simulated Data with Known Component Labels.} Each panel shows a bar chart of the average AUC scores with error bars indicating 95\% confidence intervals. Each of the five bars refers to a specific method (see legend). The rows and columns refer to sample sizes $m$ and numbers of mixture components $K$, respectively. The models clG and mBGe required inferring the data allocation from a set of binary variables.}  
\label{FIGURE_1}
 \end{center}
\end{figure}

 The clG and mBGe models are designed for hybrid data, i.e., datasets that contain both discrete and continuous variables. Both models allow edges from discrete to continuous variables. In principle, we could treat the component allocation as a single discrete variable with 
$K$ categories; however, if this variable were made a parent of all continuous nodes, clG and mBGe would effectively reduce to the full-covariance (M1) and tied-covariance (M2) mixture models with known class labels, respectively. To avoid this, we instead represent the class membership by 
$\log_2(K)$ binary indicator variables and include them alongside the continuous variables. Each value combination of the binary indicators corresponds to a specific 
$k$, and clG and mBGe are then applied in the usual way, i.e., structure learning infers which indicators serve as parents of which continuous variables. This setup is consistent with the study reported in Section~5.1 of \cite{Gryze2025}.\footnote{Note that the notation differs from the present paper: in \cite{Gryze2025} the models 
were denoted BGe (H), hBe (M1), clG, and mBGe, and were applied to learn Bayesian networks 
from hybrid data. For hybrid data, the hBe (M1) model assumes that each of the $K$ value 
combinations of the discrete variables corresponds to a mixture component with separate network 
parameters. In \cite{Gryze2025}, this assumption introduced many redundant components and thus 
weakened the performance of the hBe (M1) model. In our study, however, this assumption is 
beneficial, as it aligns with the ground truth.}\\

Figure~\ref{FIGURE_1} shows the average AUC scores for each combination of $m$ and $K$. 
Overall, the tied-covariance mixture model (M2) performs best and yields the highest AUC scores. 
H (BGe) and clG are consistently outperformed by M2. The models M1, M2, and mBGe perform 
similarly for $K=2$, but M2 achieves higher AUCs for $K \geq 4$. We also compared model 
performance in terms of relative structural Hamming distance (rSHD). The rSHD results, 
shown in Figure 1 of {\bf Part~B of the supplementary material}, are in good agreement with the AUC findings.\\

\noindent {\bf Part 2:} Second, we assume the class labels to be latent (i.e., unknown) and focus on the  performance of the two mixture GBN models, M1 (full-covariance) and M2 (tied-covariance). For details we refer to Sections~\ref{sec:mixture} and \ref{sec:mcmc}.\footnote{Latent class labels disqualify  the two hybrid data models (clG and mBGe). In the absence of discrete variables, they reduce to homogeneous  GBNs.} We also include the homogeneous Gaussian Bayesian network (H) as a reference model.  However, since H ignores the heterogeneity of the observations and merges them into a single component, its  performance does not differ from {\bf Part~1}. \\

\begin{figure}[p]
\vspace{3mm}
 \begin{center}
\includegraphics[width=0.89\linewidth]{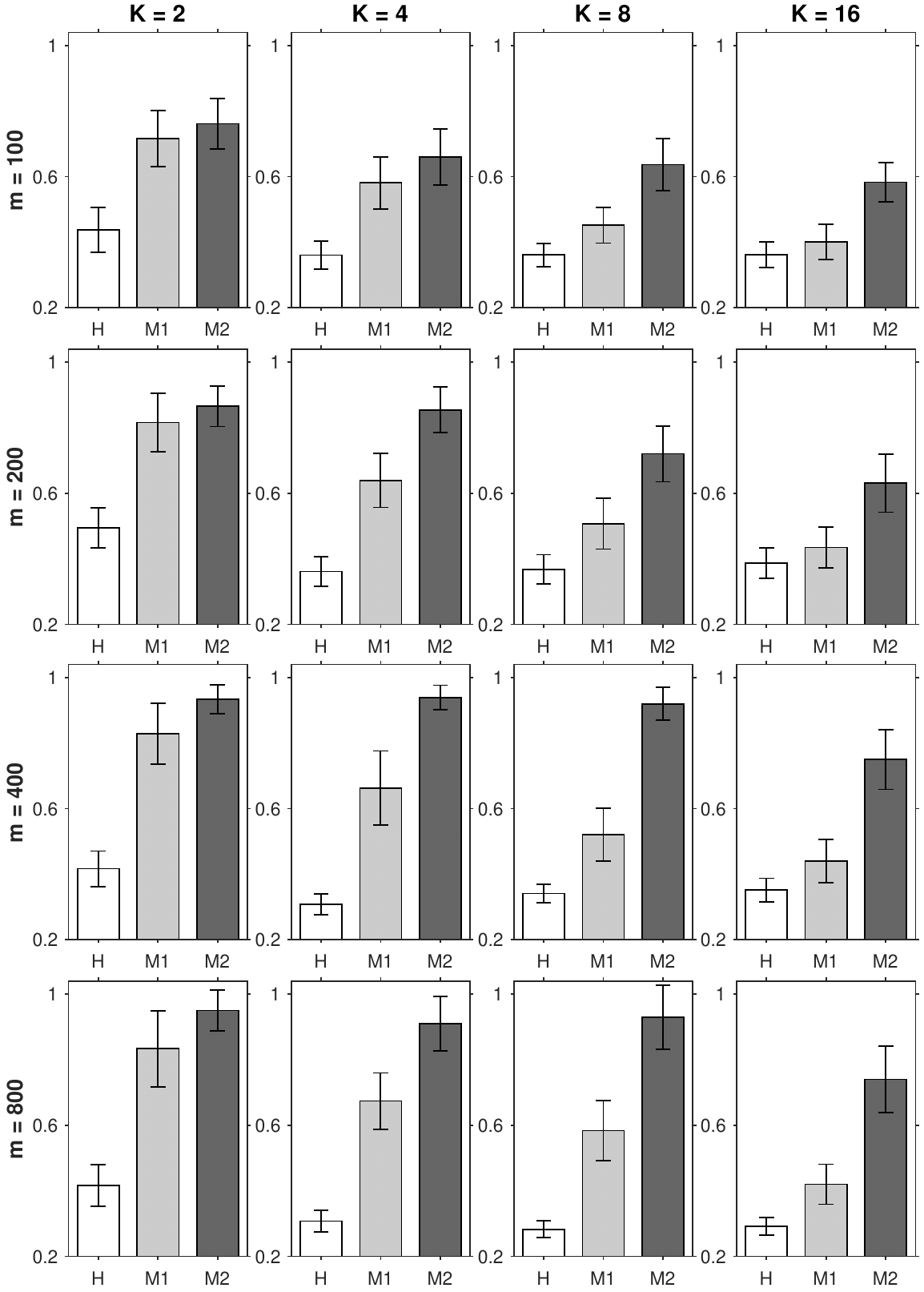} 
  \vspace{-0.35cm}
\caption {{\bf Areas under the Precision-Recall Curve (AUCs) for Simulated Data with Latent (Unknown) Component Labels.} Each panel shows a bar chart of the average AUC scores with error bars indicating 95\% confidence intervals.  The rows refer to four sample sizes $m$, and the columns refer to four numbers of mixture components $K$.}  
\label{FIGURE_2}
 \end{center}
\end{figure}

Figure~\ref{FIGURE_2} shows the method-specific average AUC scores with latent class labels. The AUC scores  of M2 decrease only slightly, indicating that the class labels can indeed be inferred from the data. Again, M2 consistently achieves the highest AUC scores. As expected, the AUC scores of both mixture models decrease  with the number of mixture components $K$, but increase with the sample size $m$. The relative structural 
Hamming distance (rSHD) results are shown in Figure~2 of {\bf Part~B of the supplementary material} and are in good agreement with the AUC findings.

\subsection{Predictive probabilities for benchmark data}
\label{sec:results2}
For the four benchmark datasets from Section~\ref{sec:data2} the true networks (CPDAGs) are unknown, 
and therefore we evaluate model performance using predictive probabilities. 
Although the class labels are known, we treat them as latent to demonstrate that they can be inferred from the data. \\

When a benchmark dataset comprises $m_{\text{all}}$ samples, we split it into a training set $\mathcal{D}$ 
and a validation set $\mathcal{D}^{\star}$. For the training set, we randomly select 
$m \in \{25,50,100,150\}$ samples from the $m_{\text{all}}$ available, and use the remaining 
$m^{\star} = m_{\text{all}} - m$ observations for validation.\footnote{The Iris dataset has only 
$m_{\text{all}}=150$ samples. We therefore restrict the maximum training size to $m=145$.} 
To ensure comparability across validation sets of different sizes $m^{\star}$, we compute 
the geometric mean predictive probabilities, as described in Section~\ref{sec:evaluation}. \\

\begin{figure}[p]
\vspace{3mm}
 \begin{center}
\includegraphics[width=0.99\linewidth]{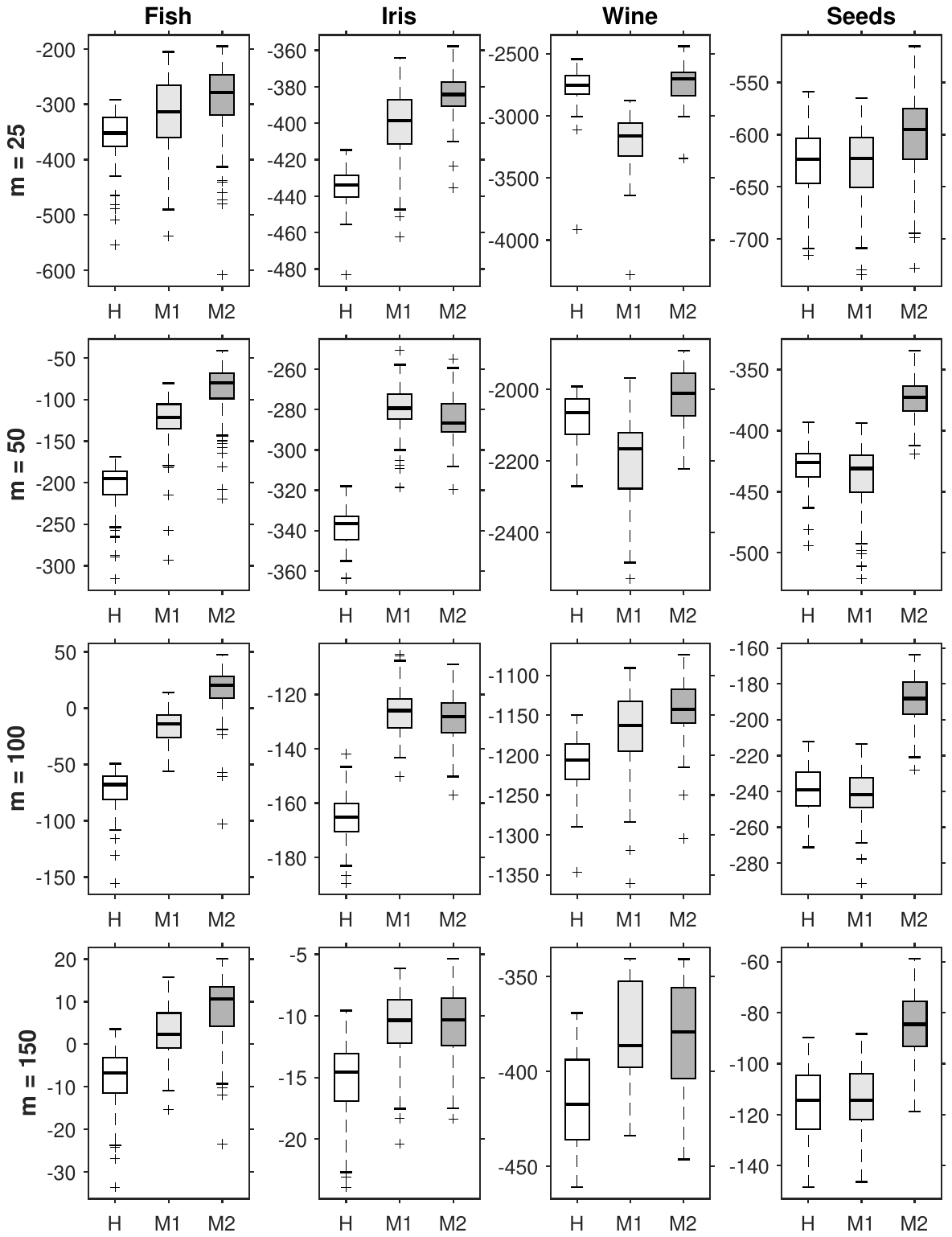} 
  \vspace{-0.35cm}
\caption {{\bf Geometric Mean Predictive Probabilities (PPs) for Benchmark Data}. Each figure shows three boxplots of the distribution of log predictive probabilities (PPs) computed from 100 random data subsets. The boxplots refer to the models H (white), M1 (light grey), and M2 (dark grey). The columns refer to four different datasets, and the rows refer to four sample sizes $m$.}  
\label{FIGURE_3}
 \end{center}
\end{figure}

Figure~\ref{FIGURE_3} shows boxplots of the geometric mean predictive probabilities. 
As expected, predictive performance improves with increasing sample size $m$, while 
relative differences between models tend to diminish. Overall, the new M2 
tied-covariance mixture GBN model yields the highest predictive probabilities, whereas 
the homogeneous GBN (H) model consistently yields the lowest. For the four benchmark datasets, we observe the following trends:
\begin{itemize}
\item {\bf Fish:} The tied-covariance mixture GBN (M2) performs best, followed by the full-covariance mixture GBN (M1), with the homogeneous GBN (H) performing worst.
\item {\bf Iris:} The homogeneous GBN (H) performs poorly. Between the mixture models, M2 is slightly better only at $m=25$; for larger $m$, their performances are similar.
\item {\bf Wine:} M2 again performs best overall. At $m=25$, the homogeneous GBN (H) is competitive, while at $m \geq 100$, M1 approaches M2’s performance.
\item {\bf Seeds:} As with the Fish data, M2 clearly outperforms the other models. Both H and M1 perform poorly, suggesting that M2 strikes the right balance between the under-complex H and the over-complex M1.
\end{itemize}

\subsection{Applications to benchmark data}
\label{sec:results3}
In Section~\ref{sec:results2}, the tied-covariance mixture GBN (M2) achieved the highest predictive probabilities overall. We therefore adopt this model for analyzing the four benchmark datasets. The purpose of this application study is twofold: (i) to demonstrate that the underlying class allocations can indeed be learned from the data, and (ii) to predict the unknown network structures (CPDAGs). Although these benchmark datasets are well known and widely used, to the best of our knowledge no network structures have been reported for them in the literature.  \\
\begin{figure}[p]
\vspace{3mm}
 \begin{center}
\includegraphics[width=0.99\linewidth]{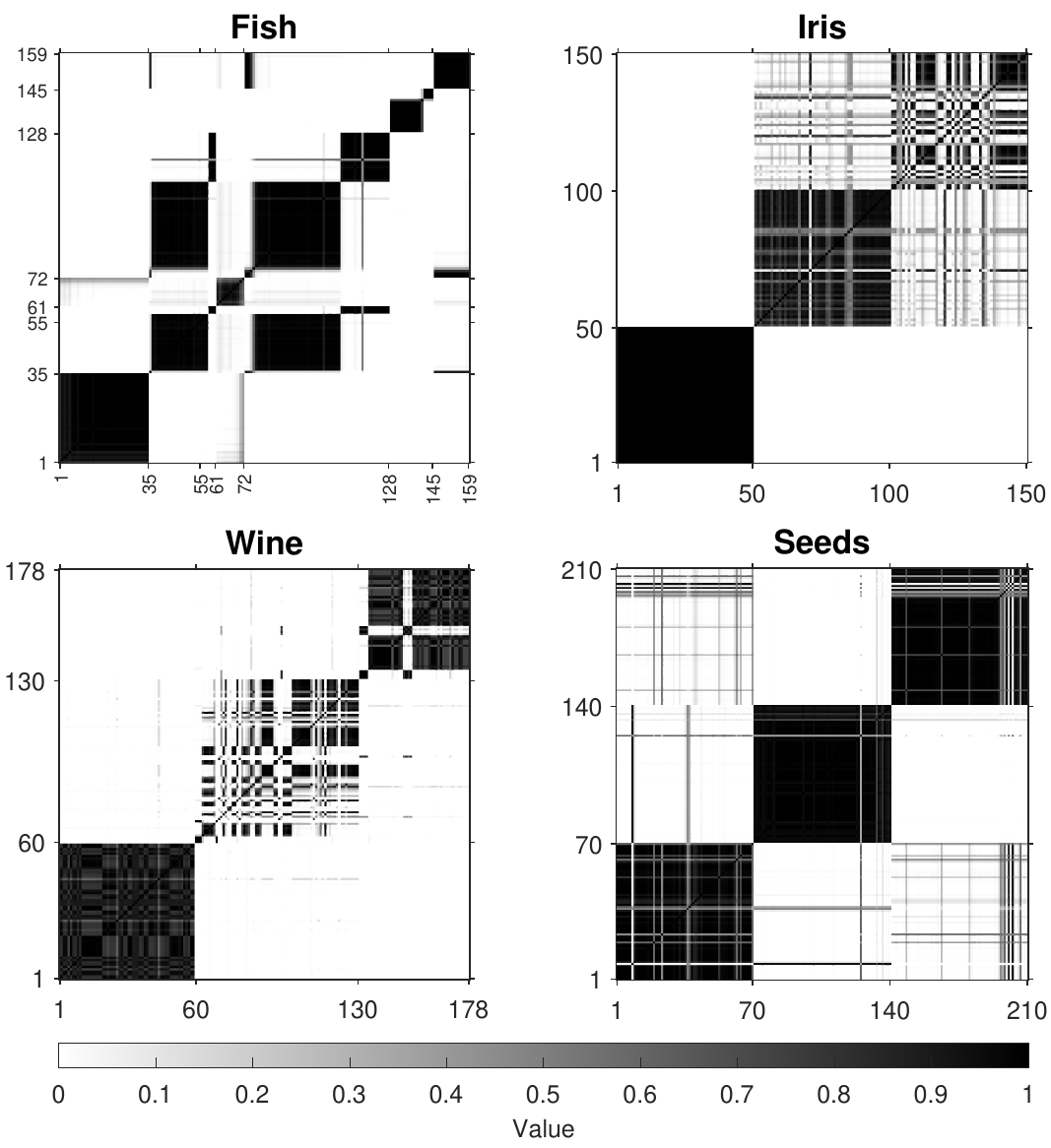} 
  \vspace{-0.35cm}
\caption{\textbf{Heatmaps of pairwise co-allocation probabilities for the four benchmark datasets.}
All panels show probabilities under the tied-covariance mixture GBN (M2) model. 
Axis tick marks indicate the positions of the true transitions between mixture components.}
\label{FIGURE_4}
 \end{center}
\end{figure}

Figure~\ref{FIGURE_4} shows heatmaps of the pairwise co-allocation probabilities. The four heatmaps reveal the following trends:
\begin{itemize}
\item {\bf Fish:} Approximately $K=7$ groups are inferred. These groups correspond broadly, but not one-to-one, to the fish species. The three largest groups are: {\em Bream} (1–35); {\em Roach} together with most of the {\em Perch} (36–55 and 76–106); and the remaining {\em Perch} with three {\em Whitefish} (107–128 and 59–61). In addition, {\em Parkki} (62–72), {\em Pike} (129–141), and {\em Smelt} together with three {\em Perch} (146–159 and 73–75) each form smaller groups.
\item {\bf Iris:} Three clear groups are inferred. {\em I. setosa} (1–50) is completely separated from the other two species, while {\em I. versicolor} and {\em I. virginica} show partial overlap. A small subset of {\em I. virginica} flowers (101–150) appears separated from the rest.
\item {\bf Wine:} Two large groups are inferred, corresponding to {\em Cultivar 1} (1–59) and much of {\em Cultivar 3} (131–178). In contrast, the wines from {\em Cultivar 2} (36–55) do not form a distinct cluster but are spread across several smaller groups.
\item {\bf Seeds:} The three wheat varieties are almost perfectly separated, indicating that the class allocation can be recovered very accurately from the data.
\end{itemize}

Overall, the co-allocation heatmaps demonstrate that the model is able to recover meaningful group structures from the data. For Seeds and Iris the inferred groups align almost perfectly with the true classes, while for Fish and Wine the model uncovers finer substructures beyond the original labels, highlighting additional heterogeneity. These results confirm that the underlying data allocation can indeed be learned directly from the data.\\

\begin{figure}[t]
 \begin{center}
\includegraphics[width=0.69\linewidth]{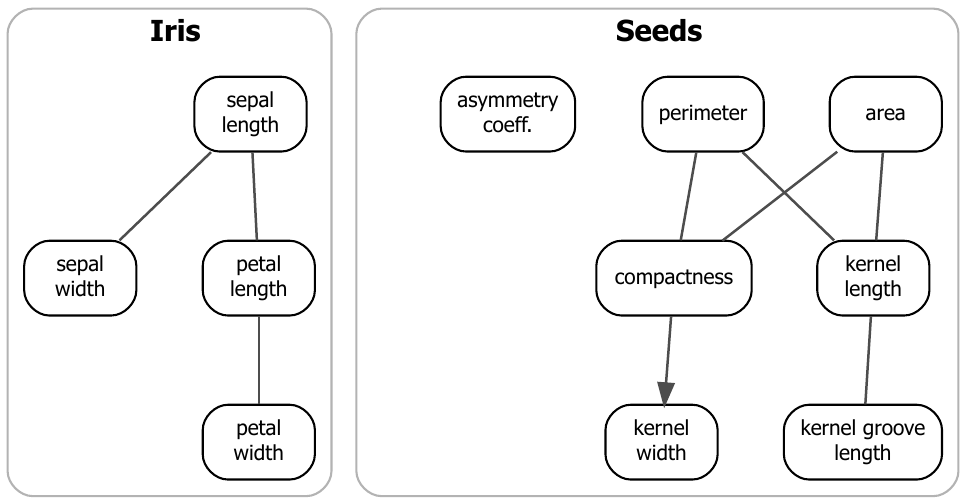} 
  \vspace{-0.35cm}
\caption{\textbf{Inferred networks (CPDAGs) for the Iris and Seeds data.} The class labels were assumed to be unknown and the tied-covariance mixture model (M2) was applied to all data. Shown are the edges whose scores exceeded the threshold $\psi=0.75$. }
\label{FIGURE_5}
 \end{center}
\end{figure}

\begin{figure}[b]
 \begin{center}
\includegraphics[width=\linewidth]{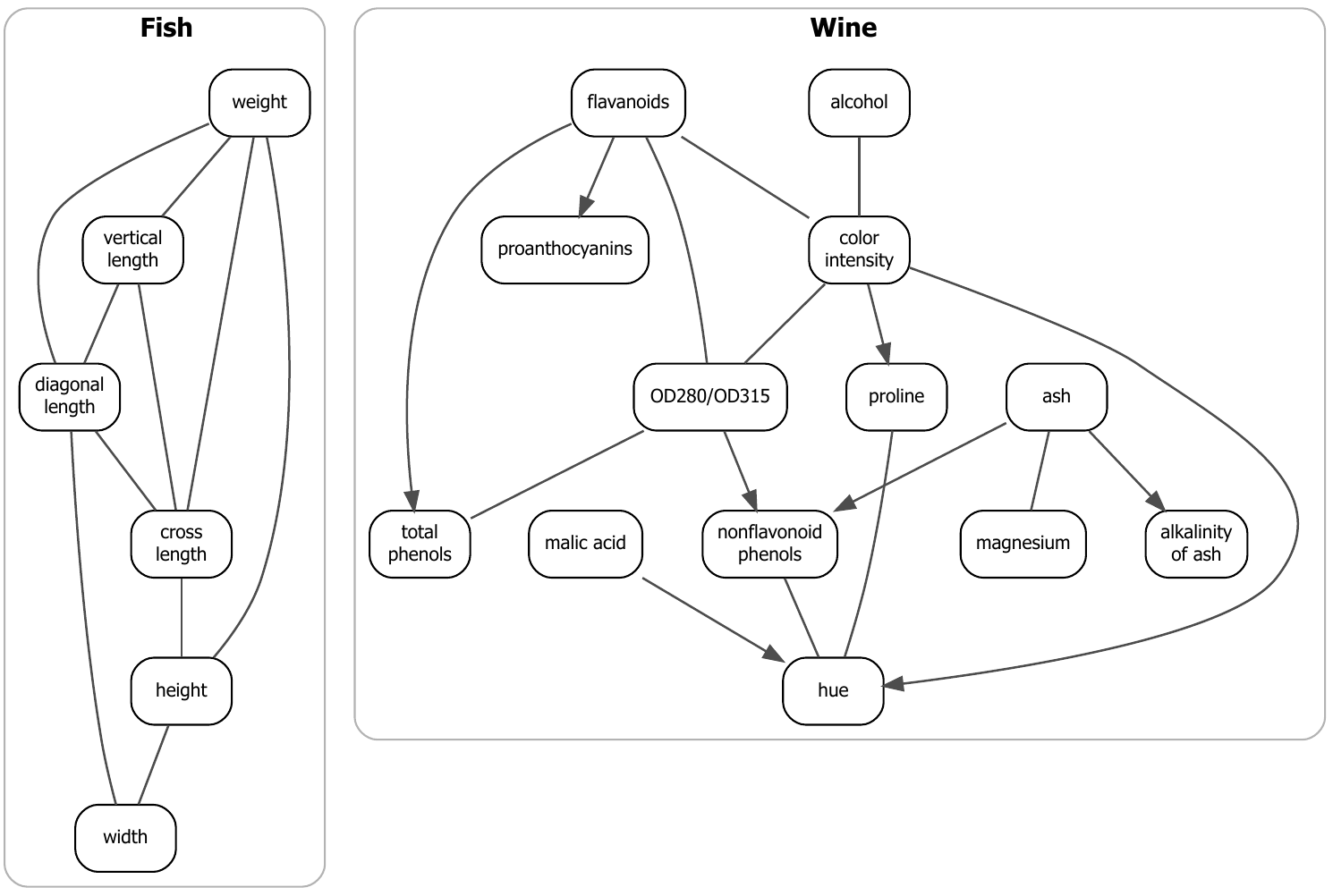} 
  \vspace{-0.35cm}
\caption{\textbf{Inferred networks (CPDAGs) for the Fish and Wine data.} The class labels were assumed to be unknown and the new tied-covariance mixture model (M2) was applied to all data. Shown are the edges whose scores exceeded the threshold $\psi=0.75$ ({\bf Fish}) and $\psi=0.5$ ({\bf Wine}), respectively.}
\label{FIGURE_6}
 \end{center}
\end{figure}

Figures~\ref{FIGURE_5}–\ref{FIGURE_6} show the four inferred networks (CPDAGs). For the {\bf Iris}, {\bf Fish}, and {\bf Seeds} datasets, we display all edges with posterior scores above the threshold $\psi=0.75$. For {\bf Wine}, however, we relaxed the threshold to $\psi=0.5$, since otherwise the network would have contained only four edges. We represent an undirected edge $X_i - X_j$ whenever one orientation ($X_i \rightarrow X_j$ or $X_j \rightarrow X_i$) exceeded the threshold $\psi$, and the opposite orientation had a comparably high score.\footnote{\scriptsize We considered $X_i \rightarrow X_j$ and $X_j \rightarrow X_i$ to have similar scores $e_{i,j}$ and $e_{j,i}$ if $|e_{i,j} - e_{j,i}| < 0.25$.}
Unlike the three smaller networks ({\bf Iris}, {\bf Seeds}, {\bf Fish}) with $n \in \{4,6,7\}$ nodes, the {\bf Wine} network with $n=13$ nodes contains many directed edges. A detailed substantive interpretation of the inferred structures is beyond the scope of this paper. Nevertheless, we hope that these exemplary applications may encourage the use of these four widely known datasets also as benchmarks for graphical model analysis (cf. Section~\ref{sec:conclusion}).

\section{Conclusion and discussion}
\label{sec:conclusion}
In this paper, we proposed an extension of the BGe scoring metric originally introduced by Geiger and Heckerman. For mixture data consisting of distinct components, the standard extension of the BGe score assumes that each mixture component has its own network parameters, namely a separate mean vector and covariance matrix. In contrast, our extension assumes component-specific mean vectors while allowing the covariance matrix to be shared across components. This construction is directly analogous to classical mixture models with full-covariance and tied-covariance, and we therefore referred to the two mixture GBNs as the full-covariance mixture model (M1) and the tied-covariance mixture model (M2).\\
Our empirical results on both simulated and benchmark data demonstrate that the M2 mixture model can yield significant improvements over the M1 mixture model. In particular, consistent with classical findings for mixture models in statistics, we observed that the tied-covariance model (M2) performs especially well at smaller sample sizes, while this advantage diminishes or may even reverse as sample sizes increase. These results underline the practical value of the new scoring metric as a complementary alternative to the standard full-covariance approach (M1).\\
We have applied both mixture models to four well-known and widely used datasets (Fish, Iris, Wine, Seeds). Although these datasets have traditionally been employed primarily to evaluate classification accuracy, we believe their heterogeneity offers considerable potential for graphical modeling. We therefore hope that our study contributes to establishing them as datasets for Bayesian network analyses. While our focus was on mixtures of Gaussian Bayesian networks, these datasets also appear well suited for assessing other Bayesian network approaches, particularly models designed to analyze heterogeneous data arising from different experimental conditions, as, for example, proposed by \cite{Scutari2022}.
\clearpage

\clearpage

\bibliographystyle{elsarticle-num-names} 
\bibliography{tdh}

\end{document}